\documentclass[aps,twocolumn,groupedaddress,showpacs,superscriptaddress,amssymb,amsmath,longbibliography]{revtex4-2}
\usepackage{graphicx}
\usepackage{amsmath}
\usepackage{amssymb}
\usepackage{amsfonts}
\usepackage{color}
\usepackage{bm}        
\usepackage{comment}
\usepackage{placeins}
\usepackage[mathscr]{eucal}
\usepackage[colorlinks, linkcolor=red,citecolor=blue,urlcolor=blue]{hyperref}
\usepackage[normalem]{ulem}
\usepackage[all]{hypcap}
\usepackage{multirow}
\usepackage[dvipsnames,usenames,table]{xcolor}
\usepackage{dcolumn}
\usepackage{hyperref}
\usepackage{bm}
\usepackage{epsf}
\usepackage{amsfonts}
\usepackage{epstopdf}%
\newcommand{\be}{\begin{equation}}
\newcommand{\ee}{\end{equation}}
\newcommand{\bea}{\begin{eqnarray}}
\newcommand{\eea}{\end{eqnarray}}

\usepackage[font=small,labelfont=bf,tableposition=top]{caption}
\usepackage[font=footnotesize]{subfig}
\usepackage[demo,abs]{overpic}

\def\be{\begin{equation}}
\def\ee{\end{equation}}
\def\bea{\begin{eqnarray}}
\def\eea{\end{eqnarray}}

\newcommand{\ket}[1]{\vert #1 \rangle}

\newcommand{\braket}[3]{\left\langle #1 \middle| #2 \middle| #3 \right\rangle}

\begin{document}

\title{Variational Quantum Eigensolvers in the Era of Distributed Quantum Computers} 

\author{Ilia Khait}
\affiliation{Entangled Networks Ltd., Toronto, Ontario, M4R 2E4, Canada}
\affiliation{Department of Physics and Centre for Quantum Information and Quantum Control, University of Toronto, Toronto, Ontario, Canada M5S 1A7}
\email{ilia@entanglednetworks.com}
    \author{Edwin Tham}
\affiliation{Entangled Networks Ltd., Toronto, Ontario, M4R 2E4, Canada.}
\author{Dvira Segal}
\affiliation{Department of Chemistry, 
University of Toronto, 80 Saint George St., Toronto, Ontario, M5S 3H6, Canada}
\affiliation{Department of Physics and Centre for Quantum Information and Quantum Control, University of Toronto, Toronto, Ontario, Canada M5S 1A7}
\author{Aharon Brodutch}
\affiliation{Entangled Networks Ltd., Toronto, Ontario, M4R 2E4, Canada}
\date{\today}

\begin{abstract}
The computational power of a quantum computer is limited by the number of qubits available for information processing. Increasing this number within a single device is difficult; it is widely accepted that distributed modular architectures are the solution to large scale quantum computing.
The major challenge in implementing such architectures is the need to exchange quantum information between modules.
In this work, we show that a distributed quantum computing architecture with {\it limited} capacity to 
exchange information between modules can accurately solve quantum computational problems.
Using the example of a variational quantum eignesolver with an ansatz designed for a two-module (dual-core) architecture, we show that three inter-module operations provide a significant advantage over no inter-module (or serially executed) operations. These results provide a strong indication that near-term {\it modular} quantum processors can be an effective alternative to their monolithic counterparts. 
\end{abstract}

\maketitle 

{\it Introduction.--} Quantum computers promise significant speed-up for a diverse set of problems~\cite{brassard2002-qaa,grover1996,peruzzo2014-vqe,farhi2014-qaoa}. 
However, the quantum advantage over classical computation only becomes appreciable when the problem size (i.e., the number of qubits required to solve the problem) is sufficiently large.
Yet in practice, increasing the number of useful qubits on a quantum processing unit (QPU) is challenging: Generally, there is a trade-off between qubit count and qubit quality~\cite{LaFlamme1996-decoherencebound,Steane2000-GateSpeed,Monroe2013-ScalingIonTraps,Murali2020}.
Modular architectures, where small high quality QPUs are interconnected, offer a more sustainable solution to the scaling problem than a monolithic approach~\cite{Monroe2014-DistributedQPU,Gold2021-RigettiInterconnect,Nickerson2014-ModularCells,Brown2016-ModularIonTraps,Bravyi2022-ScalingSuperconductingQPU, Awschalom_2021}. 
In small devices, high-fidelity qubit operations are easier to engineer, and corresponding verification and validation are more tractable.
Modular approaches, however, require transmission of quantum information between QPUs. This information exchange can be used to create effective interactions between qubits residing on different QPUs. 
In general, information transfer between different modules is significantly slower and less reliable than between qubits assigned to the same module. We call this the \emph{quantum interconnect bottleneck} (QIB). 

An increasingly salient architectural question for quantum computers concerns trade-offs in using an interconnected multi-module quantum device: Do the overheads associated with the QIB outweigh the benefits of adding qubits to a monolithic device? 

\begin{figure}[!ht]
    \subfloat{%
      \begin{overpic}[width=0.45\textwidth]{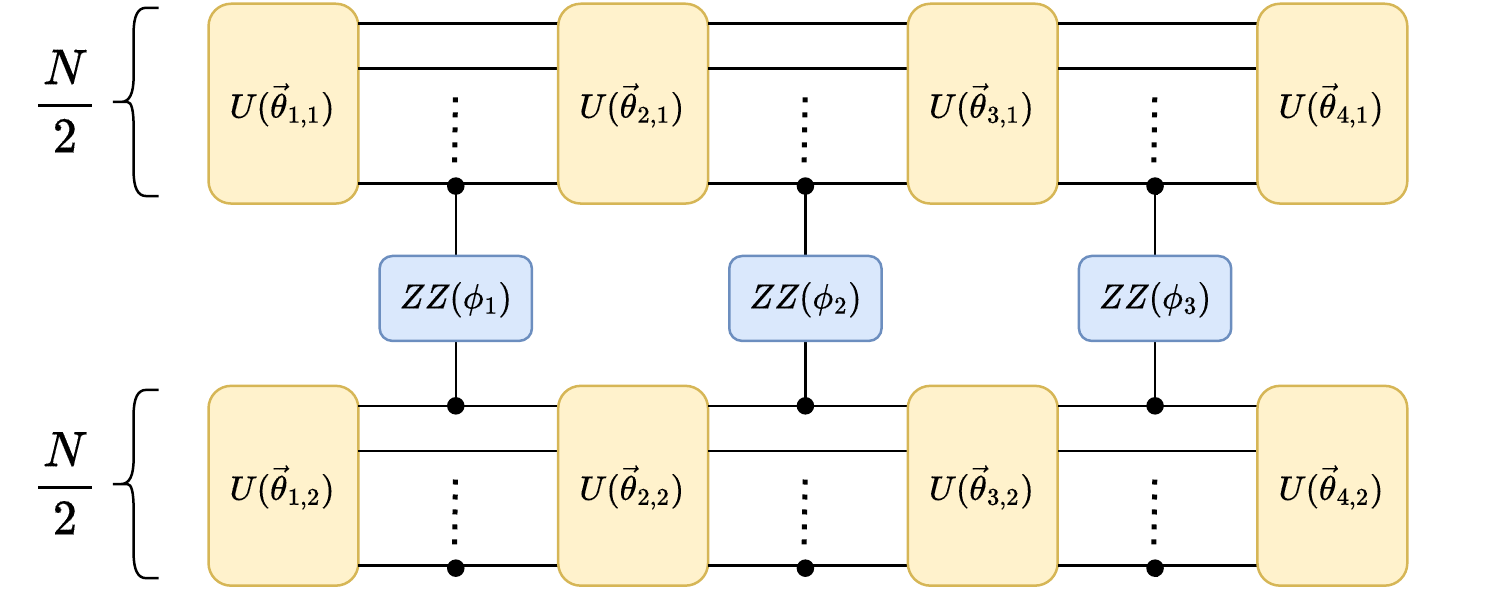}
        \put(-5,60){(a)}
        \put(-5,-45){(b)}
      \end{overpic}
    }\hfill
    \subfloat{%
      \begin{overpic}[width=0.35\textwidth]{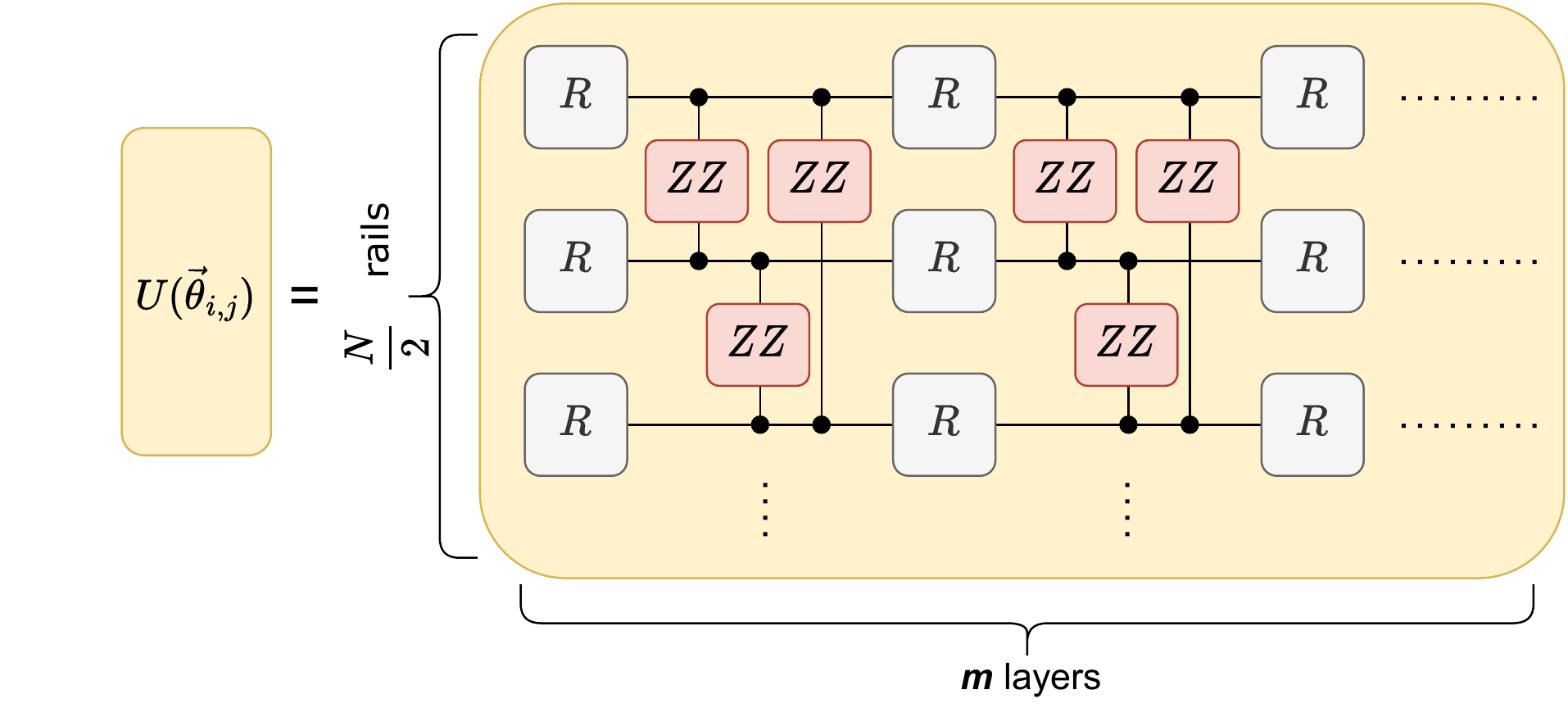}
      \end{overpic}
    }
\caption{
A VQE ansatz circuit for a dual core quantum architecture.
(a) VQE ansatz's structure: Each QPU contains half of the available qubits, $N \over 2$.  
The unitary $U(\vec{\theta}_{i,j})$   
(yellow block) acts on
the qubits of the $j$th QPU at the $i$th stage,  
 followed by a parametric remote gate operation $ZZ$ (blue block), 
 allowing entanglement to be shared between QPUs.
Throughout this paper,
the number of inter-core operations is $n_i=3$. 
(b) The  unitary $U(\vec{\theta}_{i,j})$ contains $m$ layers; gates $R$ are implicitly parameterized by elements of $\vec{\theta}_{i,j}$. Throughout this paper, $m=3$ for a total $12$ layers executed on each QPU. 
}\label{fig:VQEansatz}
  \end{figure}

Suppose  one aims to run a circuit that requires $N$ qubits but only has access to $M$-qubit devices with $M<N$. Assuming these devices can exchange quantum information using a quantum interconnect, it is possible to recompile the circuit~\cite{ENcomp} such that it  uses the interconnect $n_i$ times. 
To quantify the benefits of a quantum interconnect we compare a dual-core solution to a naïve approach with comparable running time -- solving different parts of the problem on separate QPUs, and relying solely on classical communication; we refer to this as the \emph{separable} (or $n_i=0$) solution. The dual core solution consists of two interconnected QPUs with $N \over 2$ qubits each, while assuming that each QPU individually has an all-to-all connectivity map, i.e., within each module, qubits can interact directly with every other qubit. 

If one allows $\mathcal{O}(N)$ interconnect uses, the aforementioned architecture becomes equivalent to an all-to-all $N$-qubit device.
However, the QIB combined with practical considerations, such as decoherence, requires limiting $n_i$.
As described below, 
$n_i=3$ is not only sufficient for the problems we consider, but it also shows a significant improvement over the separable solution.
Specifically, we show that for a dual-core architecture, the estimation error arising from the expressibility of a limited-connectivity ansatz is exponentially suppressed with $n_i$.

In Fig.~\ref{fig:VQEansatz}~(a) we show our variational ansatz, which is composed of single-qubit operations along with the $ZZ$-gate, 
$ZZ\left(\phi\right)=\exp\left(i\frac{\phi}{2}\sigma_{z}\otimes\sigma_{z}\right)$,
a common entangling operation in trapped-ion devices~\cite{Molmer1999-XXgate,Molmer2000-XXgateFidelity}.
We treat $ZZ$ gates that straddle two clusters of $N/2$ qubits as interconnect mediated remote gates.
We compare performance between the separable and modular architectures for a common algorithm, the variational quantum eigensolver (VQE) ~\cite{Peruzzo_2014,Fedorov2022}, which estimates the ground state energy of a Hamiltonian.
To make a simple comparison between the separable solution and the interconnected one, we only consider the precision of the result on the final circuit, and avoid issues related to the performance of the optimization stage. 

{\it Interconnect advantage.--} 
Decomposing a state into its principal components~\footnote{This decomposition can be achieved via SVD, or the Schmidt decomposition, which is used here.} is a core technique in many numerical recipes such as the DMRG~\cite{PhysRevLett.69.2863}, where a truncation is performed based on the diminishing return on fidelity of storing more basis components (at a high cost). Such a rationale is used for understanding the power of interconnects. 
Suppose  one has two QPUs, each capable of preparing any state in its $M$-qubit Hilbert space. The state of that dual-QPU system is a product state $\ket{\psi} = \ket{\psi_1}\ket{\psi_2}$, where $\ket{\psi_i}$ is the state of the $i$th QPU.
Every application of a remote operation between QPUs increases inter-QPU entanglement, as expressed by the rank ($d$) of the Schmidt decomposition of $\ket{\psi}$, cut along the two QPUs: $\psi^{(d)} = \sum_{i=1}^d c_i \ket{\psi_1^{(i)}}\ket{\psi_2^{(i)}}$.
Note that with a sufficiently expressive \textit{intra}-QPU ansatz ($U(\vec{\theta})$ in Fig.~\ref{fig:VQEansatz}) the entanglement rank $d$ can rise quickly, up to exponential in number of inter-QPU operations $n_i$ (i.e. $d\leq 2^{n_i}$).

{\it Procedure.--} 
VQE is an iterative classical-quantum hybrid algorithm, which estimates the ground state energy of a given Hamiltonian~\cite{peruzzo2014-vqe,Kandala2017}.
A quantum computer produces an approximation of the Hamiltonian's ground-state based on a parameterized ansatz; in turn, a classical strategy for converting a Hamiltonian into a series of compactly-implementable observables estimates an energy-eigenvalue from that approximate ground-state~\cite{Romero_2019,Ryabinkin2018,Ryabinkin2020}.
This process is repeated, with varied ansatz parameters chosen by an optimization strategy until a sufficiently refined ground-state is reached~\cite{Moll_2018}.
In what follows, we demonstrate that the $n_i=3$ dual-core parameterized circuit suggested in Fig.~\ref{fig:VQEansatz} 
provides an excellent approximation to the exact ground state of interacting systems.

Apart from ansatz expressibility (how closely the ansatz can approximate an arbitrary quantum state), VQE results also depend on classical factors like the Hamiltonian-to-observable map and parameter optimization routines.
Since we intend to test how expressibility is augmented by interconnects in a multi-QPU setup, we avoid confounding classical issues by maximizing the fidelity between the variational state and the exact ground state, instead of minimizing the expectation value of the Hamiltonian, $E_{\rm var}$. 
The following summarizes our procedure; for details  see Ref. \cite{SI}.
First, we diagonalize the Hamiltonian obtaining the exact ground state, $\ket{\psi_{\rm GS}}$. We perform an SVD (where the system is divided into two units), 
and retain the eight most significant contributions,
\bea
\ket{\psi_{\rm GS}} \rightarrow \ket{\psi_{\rm GS}^{(8)}} =  \sum_{i=1}^{8} \lambda_i \ket{\phi_1^{(i)}}\ket{\phi_2^{(i)}}.
\label{eq:SGS}
\eea
Here, $\lambda_i$, $\ket{\phi_1^{(i)}}\ket{\phi_2^{(i)}}$ are the $i$th Schmidt eigenvalue and eigenvector, respectively.
We iteratively build the variational solution, 
(see Fig.~\ref{fig:VQEansatz}): 
We start by optimizing the fidelity towards a product state, our
 crudest approximation to the ground state, $\ket{\psi_{\rm GS}^{(1)}}$,  
defined by the largest Schmidt eigenvalue $\lambda_1$. We construct a variational approximation of this target state using the first set of unitaries $U(\vec{\theta}_{1,1})$ and $U(\vec{\theta}_{1,2})$, by applying them on an all-polarized state $\ket{0 \dots 0}_1\ket{0 \dots 0}_2$. 
Next, we add another Schmidt coefficient and target the state $\ket{\psi_{\rm GS}^{(2)}}$ by enlarging the set of variational parameters: A first remote operation $ZZ(\phi_1)$ and another set of unitaries, $U(\vec{\theta}_{2,1})$ and $U(\vec{\theta}_{2,2})$. 
After optimization, we add another Schmidt coefficient, with another interconnected-remote operation, and repeat till $\ket{\psi_{\rm GS}^{(8)}}$~\footnote{Note that this is not a viable VQE optimization method in a ``production'' use since knowledge of the exact ground-state cannot be assumed; we employ it here only to test expressiblity.}.

\begin{figure}
    \includegraphics[width=0.8\columnwidth]{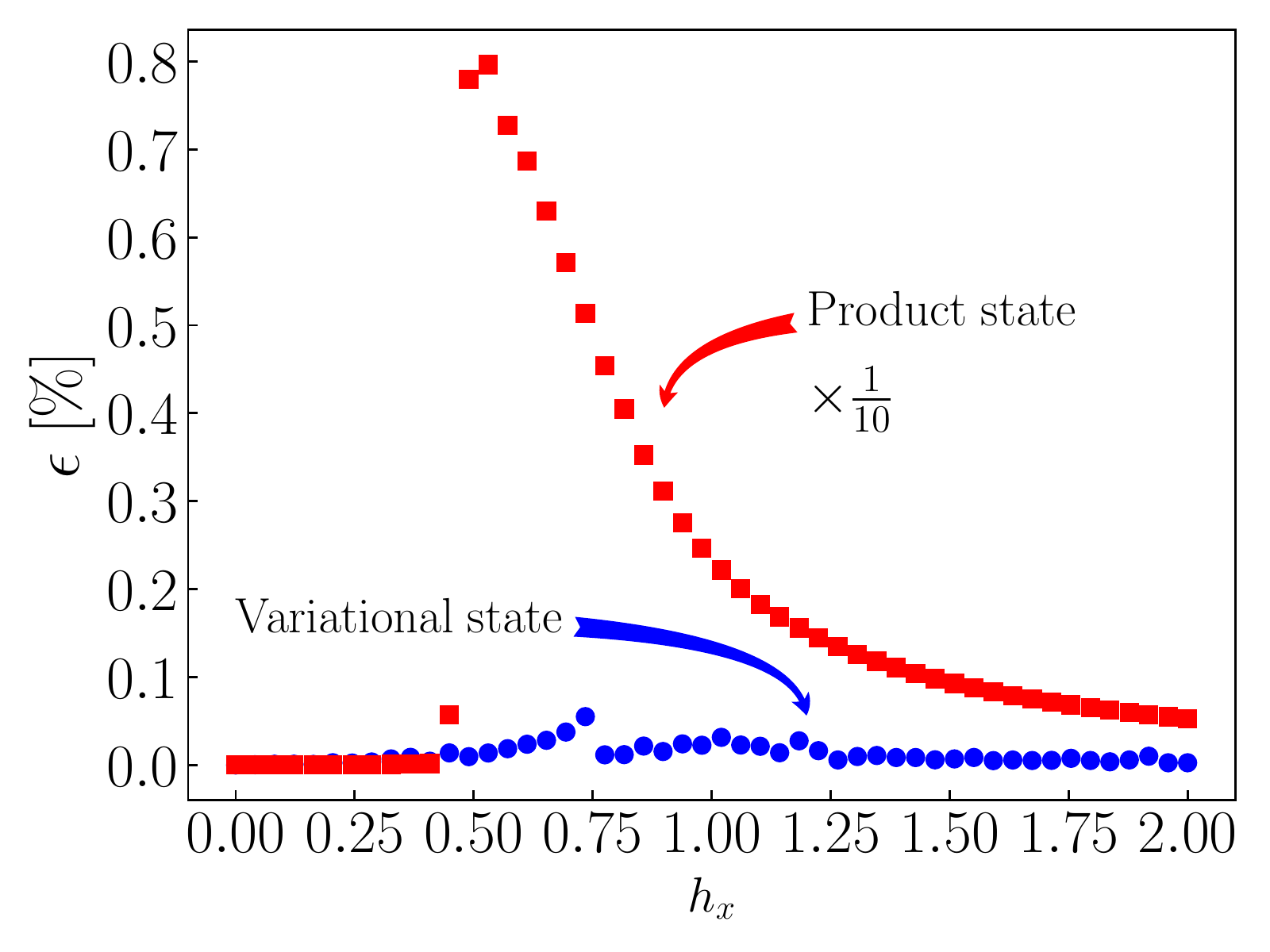} 
    \vspace{-3mm}
    \caption{TFIM model with 12 spins at $J=1$,  Eq.~(\ref{eq:TFIM}), 
studied as a function of the transverse magnetic field $h_x$. We depict (in percents) the relative energy difference between the \emph{exact ground state} and the variational ansatz, $\epsilon$. 
The two variational ansätze are a product state (red squares), in which a separable solution is forced, and the interconnected solution (blue circles). 
For presentation purposes, the error of the separable solution is decreased by an order of magnitude. 
For either vanishing or strong magnetic fields, the exact solution is a product state. However, for intermediate  $h_x$ values the separable solution is far inferior. 
}
    \label{fig:TFIM}
\end{figure}


\begin{figure*}
    \begin{overpic}[width=0.8\textwidth]{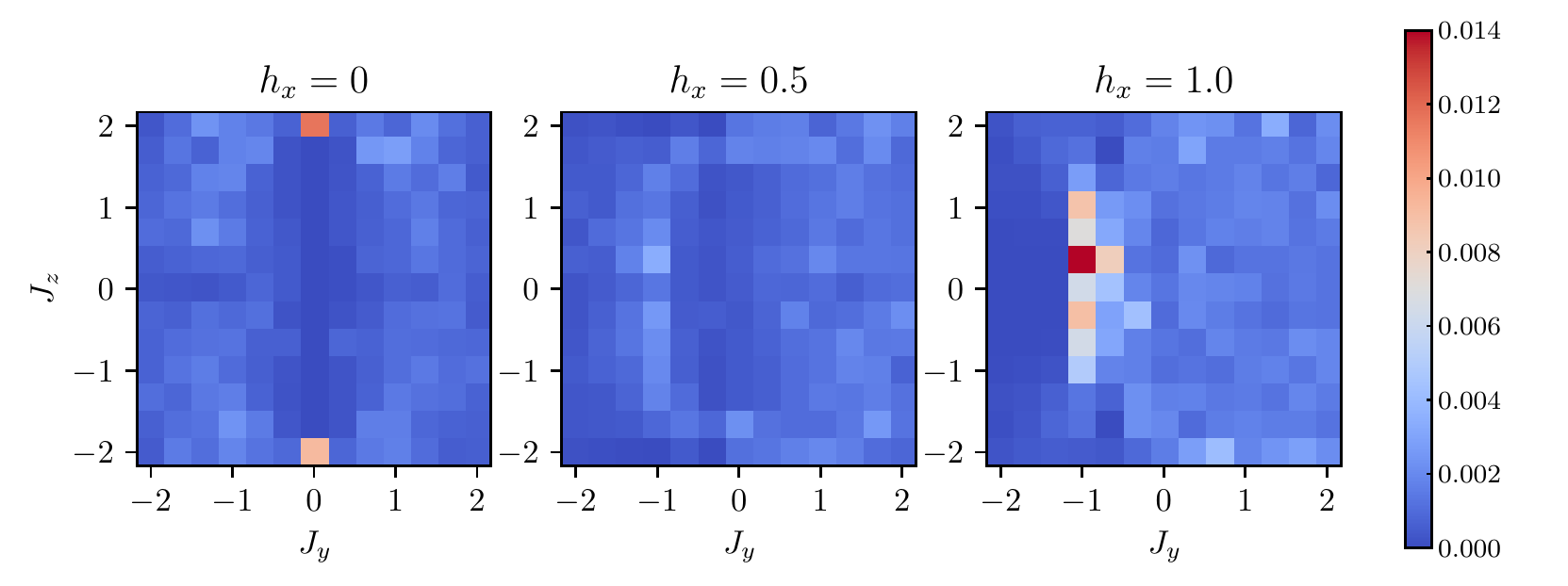} 
        \put(35,140){(a)}
        \put(145,140){(b)}
        \put(255,140){(c)}
        \put(367,150){$\epsilon$}
    \end{overpic}
    \vspace{-3mm}
    \caption{VQE results for the one-dimensional $XYZ$ model with 12 spins, see Eq.~(\ref{eq:XYZ}). We present a color map of the relative energy error $\epsilon$. The panels depict different magnetic field strengths: $h_x=0,~0.5, 1.0$, and $J_x=1$. Fidelities are usually above 99.9\%; the highest reported variation is $1.4\%$. 
    \label{fig:Heis}}
\end{figure*}


\begin{figure}
    \includegraphics[width=0.8\columnwidth]{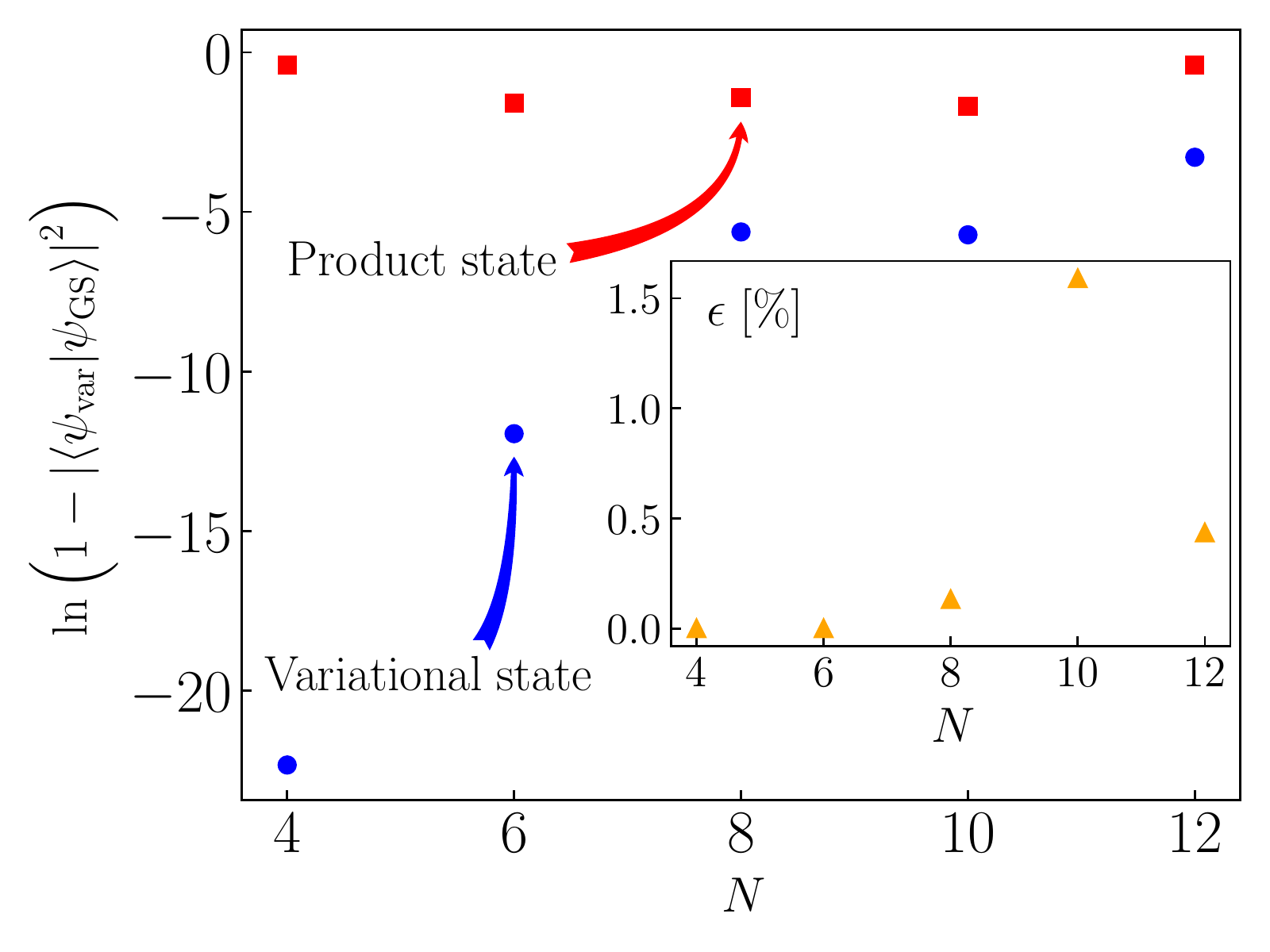} 
    \vspace{-3mm}
    \caption{VQE results for the $S$=1 Heisenberg model,  Eq.~(\ref{eq:AKLT}), 
    with up to a 6-spin chain represented by $\times 2$ qubits, $N$. 
    We display the log infidelity of the varialtional state compared to the exact ground state for the separable product states of two $N \over 2$ qubits (red squares) and the interconnected anzatse (blue circles). 
The inset shows the relative energy error
$\epsilon$, in percents. 
While the error 
of the intreconnected solution increases with $N$, it still delivers orders of magnitude better fidelity than the separable solution.
}
    \label{fig:AKLT}
\end{figure}


{\it Models.--}\label{sec:Models} We test three models to benchmark the interconnect-mediated ansatz: the transvese field Ising model (TFIM), the spin-half anisotropic Heisenberg model, and the spin-one Heisenberg model. The first two models are paradigmatic examples in benchmarking performances of novel methods~\cite{Carleo_2017,Buessen}; the spin-one Heisenberg model enables exploration of the impact of less local interactions (due to the casting onto spin-half operators) on the quality of the interconnected QPU solution.

The TFIM is defined as
\be
H_{\rm TFIM} = -J \sum_{i=1}^{N-1} \sigma_i^z \sigma_{i+1}^z -  h_x\sum_{i=1}^N \sigma_i^x,
\label{eq:TFIM}\ee
where $\sigma^\alpha,~\alpha=x,y,z$ are the Pauli matrices and $N$ is the number of spins (qubits). The phase diagram of the TFIM consists of a (anti-)ferromagnetic ordered product state for positive (negative) $J$ spin-spin interaction and vanishing transverse field $h_x$, and a disordered state at strong transverse magnetic field. In the thermodynamic limit, a quantum phase transition to a gapless phase occurs at $h_x=J$.

The anisotropic Heisenberg model, which is used in studies of magnetic systems is given by
\begin{align}
H_{{\rm XYZ}}= & \sum_{i=1}^{N-1}\left(J_{x}S_{i}^{x}S_{i+1}^{x}+J_{y}S_{i}^{y}S_{i+1}^{y}+J_{z}S_{i}^{z}S_{i+1}^{z}\right)\nonumber \\
 & +h_{x}\sum_{i=1}^NS_{i}^{x},
 \label{eq:XYZ}
\end{align}
where $S^\alpha,~\alpha=x,y,z$ are spin-half operators. 

Lastly, the $S=1$ Heisenberg model resembles the AKLT model~\cite{PhysRevLett.59.799, Smith_AKLT_2022}, and it is of great interest due to its topological properties. Here, it allows us to probe and benchmark the interconnected ansatz for a Hamiltonian with fewer local operators,
\begin{align}
H_{{\rm Heis}}= & J\sum_{\alpha,i=1}^{N-1}S_{i}^{\alpha}S_{i+1}^{\alpha}\nonumber \\
\rightarrow & \frac{J}{4}\sum_{\alpha,i=1}^{N-1}\left(\sigma_{2i-1}^{\alpha}+\sigma_{2i}^{\alpha}\right)\left(\sigma_{2i+1}^{\alpha}+\sigma_{2i+2}^{\alpha}\right)\nonumber \\
  & +J_{{\rm FM}}\sum_{\alpha,i=1}^{N}\sigma_{2i-1}^{\alpha}\sigma_{2i}^{\alpha}.
 \label{eq:AKLT}
\end{align}
Here $S^\alpha,~\alpha=x,y,z$ are spin-one operators, and the mapping 
marked by the arrow splits every $S=1$ operator into a pair of spin-half operators. $J_{\rm FM} \gg J$ is chosen such that a triplet is selected for every \emph{ other} bond (originating from the $S=1$ operators), with the resulting interactions having distance of $4$ units. 

{\it Results.--}
We study systems with up to $12$ qubits,
and demonstrate next an immense advantage for $n_i=3$ over $n_i=0$.
In Fig.~\ref{fig:TFIM}, we compare the ground state approximation for the TFIM given the two distinct architectures.
We denote the VQE solution by $E_{\rm{var}}$, and we compare it to the \emph{exact ground state} energy $E_{\rm{GS}}$.
Specifically, throughout the paper we analyze the error measure $\epsilon = \frac{E_{\rm var}}{E_{\rm GS} } -1$. 

An indication of a phase transition in the thermodynamic limit is detectable even in this small system, as can be seen in Fig.~\ref{fig:TFIM} by examining the product state (red squares). 
While for extreme field values ($h_x \rightarrow  0$, and ${h_x \over J} \gg 1$) the product state solution well approximates the exact ground state, around ${h_x \over J} \approx 0.5$ this approximation completely fails. In contrast, the $n_i=3$ ansatz maintains a lower error throughout. 
For presentation purposes, we display the product state error scaled down by a factor of $10$. Overall, the interconnected solution performs well, and its error does not exceed $0.07\%$ throughout the phase diagram, compared to an error of up to $8\%$ in the separable solution. This is comparable in magnitude to the exact (non-variational) wavefunction truncated to $8$-Schmidt terms, where infidelity is $0.01\%$.   

In Fig.~\ref{fig:Heis}, we examine a portion of the phase diagram of the anisotropic Heisenberg model 
at the fixed value $J_x=1.0$ and magnetic field values $h_x\in\{0,0.5,1.0\}$, thus including the symmetric Heisenberg point, $(J_x,J_y,J_z,h_x)=J(1,1,1,0)$. We plot $\epsilon$ for a $12$-qubit problem.  Excluding extreme malperforming data points, the energy convergence ratio stays well below $10^{-3}$~\footnote{See Ref.~\cite{SI}, Fig.~\ref{fig:ICnum} for the comparison with a separable solution.}:
The worst performing data point, occurring in the vicinity of $(J_x,J_y,J_z,h_x)=(1,-1,0.5,1)$ along the $J_z$ direction, appears to be an outlier with an error 
$\epsilon\approx 1.4\%$. We are unable to pinpoint the reason for this failing; neighboring data points in the $J_y$ direction show significantly better convergence. 

In Figure~\ref{fig:AKLT}, we study the
$S=1$-Heisenberg model at $J_{\rm FM}=10 J$~\footnote{In choosing the value $J_{\rm FM}$ we tested convergence of the resulting ground state with DMRG calculations and found complete agreements for $J_{\rm FM} > 6$.}, see Eq.~(\ref{eq:AKLT}).
The main panel presents the log-infidelity of the result, $\ln{\lbrace 1-\left|\langle \psi_{\rm var} | \psi_{\rm GS} \rangle \right|^2 \rbrace}$. 
The $n_i=3$ solution (blue circles) displays significantly better fidelities compared to the separable one ($n_i=0$, red squares), with three orders of magnitude {\it decrease} in infidelity at $N=12$, and
significantly better results for smaller systems. The inset shows the relative energy estimation error $\epsilon$ 
as a function of the number of qubits. Besides a single outstanding point ($N=10$), the upward trend reflects the increasing complexity of the solution as the system size grows.  
The outlier at $N=10$ was further examined by introducing another layer, after the third remote operation (with the VQE ansatz including 13 layers in total). This brought the relative energy error to $0.2\%$, consistent with the linear trend seen in Fig.~\ref{fig:AKLT}. Introducing the same change to other values of $N$ showed no significant change. We attribute this deviation to the optimization process as elaborated on next. 

{\it Discussion.--}
Two issues limit convergence to the exact solution: (i) Classical optimization -- As described in the Procedure section, we are using differentiable programming~\cite{JMLR:v18:17-468,schaferDifferentiableProgrammingMethod2020a} to find the optimal variational parameters for the VQE ansatz, a task of growing complexity when increasing the qubit count. The number of parameters increases depending on the ansatz structure (in our ansatz for $N=12$ 
we have $753$ variational parameters).
 (ii) Expressibility of the ansatz~\footnote{For a deeper discussion about the expressibility of parameterized quantum circuits we refer the reader to Ref.~\cite{Sim_2019}.} -- As discussed above, one can separate the effect of the interconnect (the remote gate) from the ``local'' layers (the unitaries $U(\Vec{\theta}_{i,j})$). 
 Each interconnect operation doubles the potential 
 Schmidt rank of the state,  and the role of subsequent layers is to facilitate quantum information spreading. Whether information spreads far enough depends on the number of layers and their inner structure (Fig.~\ref{fig:VQEansatz}). 
To assess the expressibility of the ansatz without
the effect of the interconnect, we have examined in Ref. \cite{SI} a single QPU architecture with all-to-all connected qubits.
While an all-to-all architecture performs slightly better than the interconnected one, it comes at a greater cost 
as increasing the number of qubits on a QPU is a non trivial task, which a multi-QPU modular architecture aims to avoid. 

The SVD eigenvalues of the TFIM decay faster than those of the $S=1$-Heisenberg model due the topological nature of the latter's ground state ~\cite{PhysRevB.48.3844}. The \emph{theoretical} lower bound on the infidelity is the sum of the discarded SVD eigenvalues squared. Considering that only three remote operations were allowed here, the discarded weight in the TFIM model was found to be $4 \cdot 10^{-10}$. 
Hence, the resulting infidelity, $10^{-6}$ 
is not limited by the interconnect. Similarly, in the $S=1$ Heisenberg with $N=12$ 
the discarded weight is $2 \cdot 10^{-3}$, 
and the reported infidelity of $3 \cdot 10^{-3}$
is close to this bound.
In conclusion, the limiting factors in solving the VQE on an interconnecetd hardware are classical optimization combined with the limited expressiblity of the ansatz, rather than the introduction of remote operations. Interestingly, we note that comparing the TFIM to the $S=1$-Heisenberg model,  the TFIM is converging much better than the latter, both in the interconnected case and in the all-to-all connected case~\cite{SI}. This could be explained by the suitability of the ansatz to the specific model; though in-depth consideration of this aspect is outside the scope of this paper.

{\it Conclusions.--} In this work, we demonstrated
that a distributed quantum architecture with only modest inter-QPU capacity
provides a dramatic advantage in VQE computations over serial architectures with no interconnects, and is on par with an all-to-all connected QPU~\cite{SI}.
In all cases studied, we found that 
three judiciously placed inter-QPU gates were sufficient to produce a significantly better approximation to the ground state energy for Hamiltonians of interest compared to an architecture with no quantum interconnects. 
For the Hamiltonians studied here we find that increasing $n_i$ allows for an exponential improvement in the fidelity~\cite{SI}.
Our comparison is based on simulations, and it is therefore limited to a small number of qubits, which allowed us to overcome some aspects of classical optimization. 
The main conclusion from this work is that an exponential increase in the Schmidt rank w.r.t. $n_i$ (and subsequently in the dimension of the effective Hilbert space)  
manifests itself when solving a practical algorithm. 

Methods such as entanglement forging and circuit knitting~\cite{Harrow2020-CircuitCut,arxiv.2205.00016,Eddins_2022} attempt to overcome the absence of entanglement at the cost of running narrow circuits more times. If the QIB is ignored, an interconnected approach is favorable since it requires exponentially fewer shots. 
However, with QIB overheads, the operation of an interconnect may extend run-times and reduce result quality.
These techniques scale exponentially worse with increased number of interconnect uses $n_i$. As such, we expect that in a future work these methods could be combined with interconnects to increase the effective Hilbert space by using classical and quantum resources.

A number of important questions remain open, including the impact of noise and the imperfect nature of interconnects. Slow interconnects would increase run-time and make the computation more susceptible to decoherence; inter-QPU operations are generally expected to have lower fidelity \cite{Stephenson2019-OpticalInterconnect,Awschalom_2021}; the use of fixed resource states creates overheads in gate-counts and the actual implementation of the interconnect would impact the other qubits. 
These limitations need to be weighed against the downside of increasing qubit count in a monolithic architecture, as well as artificially increasing qubit size using classical resources. Identifying algorithms where a limited number of interconnect uses can be proved advantageous will be an incentive for the implementation of multi-QPU architectures. We hope that our study would stimulate further work in this direction.

{\it Acknowledgments.--}  
The authors acknowledge fruitful discussions with Finn Lasse Buessen and Kevin Smith. The work of IK was supported by the Centre for Quantum Information and Quantum Control (CQIQC) at the the University of Toronto.
DS acknowledges support from an NSERC Discovery Grant and the Canada Research Chair program.


%

\newpage
{\Large Supplementary Information File}
\renewcommand{\thefigure}{S\arabic{figure}} 
\renewcommand{\thesection}{S\arabic{section}} 
   \renewcommand{\theequation}{S\arabic{equation}}
\setcounter{equation}{0}  
\setcounter{section}{0} 
\setcounter{figure}{0} 

\section{Optimization method}
As the goal of this work is to test effects of a distributed ansätze, we eschew execution on a real hardware in favor of simulators. This is done for three reasons:
\begin{itemize}
\item The hardware we envision has not yet been demonstrated in full. 
\item We want to decouple the interconnect bottleneck from other performance issues (e.g., gate imprefections).
\item We wish to avoid issues related to shot-noise.
\end{itemize} 
While evidently not scalable to an arbitrary number of qubits, direct simulations of distributed ansätze give us access to exact statevectors and energy eigenvalues for problem instances of modest size.  
This helps us circumvent uncertainties in the objective function that are inherently problematic for variational algorithms. Even under ideal hardware, shot noise drops slowly w.r.t. the number of shots ($\mathcal{O}(1/\sqrt{N})$). 
Furthermore, in order to mitigate improper optimization of variational parameters as a confounding factor in our findings, we implement here a statevector simulator under the JAX auto-differentiation framework so that training can leverage a more efficient gradient-based optimizer like Adam~\cite{pqsim, jax2018github, Kingma2014-AdamOptimizer}.

On sufficiently small problem instances, direct diagonalization of Hamiltonians of interest remains tractable. We leverage this, along with the fact that truncation by Schmidt rank is possible after each use of an inter-QPU ZZ-interaction. 
Let us denote the exact ground eigenstate of our Hamiltonian, obtained through direct diagonalization, as $\ket{\psi_{\rm GS}}$. Through application of SVD, we can write it as
\bea
\ket{\psi_{\rm GS}} \rightarrow \sum_{i=1}^d \lambda_i \ket{\phi_1^{(i)}}\ket{\phi_2^{(i)}},
\eea
where $\lambda_i$, $\ket{\phi_1^{(i)}}\ket{\phi_2^{(i)}}$ are the $i$-th Schmidt eigenvalue and eigenvector respectively, ordered by decreasing eigenvalue magnitude (i.e., with $\lambda_1$ being the largest).
Further, $\ket{\phi_1}$ and $\ket{\phi_2}$ each resides within respective halves of the QPU comprising clusters of $N/2$ qubits (see Fig.~\ref{fig:VQEansatz}).
If we truncate the ansätze shown in Fig.~\ref{fig:VQEansatz} \emph{before} the first inter-QPU gate (denoted $ZZ(\phi_1)$), then one should expect a product state consisting of \emph{exactly one} Schmidt term. We therefore train such a truncated ansätze towards the leading Schmidt term of $\ket{\psi_{\rm GS}}$, $\ket{\phi_1^{(1)}}\ket{\phi_2^{(1)}}$.
Upon adding more layers to the ansätze, with every additional inter-QPU operation we train towards a new target state augmented by additional Schmidt terms for $\ket{\psi_{\rm GS}}$.
Thus, we iteratively build the variational solution, one ansatz layer at a time, eventually capturing all non-trivial Schmidt terms of the ground-state. The motivation for such an approach is reminiscent of other layer-by-layer training approaches to variational algorithms, where training errors in challenging cost-function landscapes are mitigated by a reduction in the number of independent variational parameters within each layer~\cite{Cong2019-QuantumCNN,Babush2021-VariationalAlgos,Coles2021-CorrelatedVariationalAlgo}.

The cost-function we elected to use after each ansatz layer is the fidelity, $\cal{F}$, between some target state $\ket{\psi_{\rm targ}}$ and the state produced by the variational ansätze $\ket{\psi_{\rm var}}$. Note that this fidelity is available to us because we simulate the action of the ansätze as well as diagonalize the desired Hamiltonian directly. A more general ``in-the-field'' use of VQE will require more careful selection of a cost-function that is practically accessible.

For clarity, we provide next a step-by-step description of our training procedure. 
Starting with the ansätze (see Fig.~\ref{fig:VQEansatz}) up to but excluding the first inter-QPU gate ($ZZ(\phi_1)$), we train the unitaries by setting $\ket{\psi_{\rm targ}}=\ket{\psi_{\rm GS}^{(1)}}=\ket{\phi_1^{(1)}}\ket{\phi_2^{(1)}}$. 
We vary the variational parameters to maximize the fidelity:
${\cal F} = \left| \langle \psi_{\rm PS} | \psi_{\rm var} \rangle \right|^2  =  \left| \braket{\phi_1^{(1)}} {U ( \Vec{\theta}_{(1,1)} )} {0 \dots 0}_1 \right|^2 \cdot \left| \braket{\phi_2^{(1)}} {U ( \Vec{\theta}_{(1,2)} )} {0 \dots 0}_2 \right|^2 $,
with $\ket{\psi_{\rm PS}} =  \ket{\phi_1^{(1)}}\ket{\phi_2^{(1)}}$ the product state defined above, and $\ket{\psi_{\rm var}}$ as the variational state.

Next, we extend the ansätze to include all layers and parameters up to but excluding the 2nd inter-QPU gate $ZZ(\phi_2)$, and train the circuit towards
\[
\ket{\psi_{{\rm targ}}}=\frac{\lambda_{1}\ket{\phi_{1}^{(1)}}\ket{\phi_{2}^{(1)}}+\lambda_{2}\ket{\phi_{1}^{(2)}}\ket{\phi_{2}^{(2)}}}{\sqrt{\lambda_{1}^{2}+\lambda_{2}^{2}}},
\]
 Existing variational parameters, i.e., $\vec{\theta}_{(1,1)}$ and $\vec{\theta}_{(1,2)}$ are initialized to their pre-trained values from the previous iteration, and new parameters ($\vec{\theta}_{(2,1)}$ and $\vec{\theta}_{(2,2)}$) are initialized randomly. This procedure is simply repeated, with the inclusion of additional Schmidt terms (up to $d=$ terms) in $\ket{\psi_{\rm targ}}$ in lock-step
with expansion of the ansätze to contain additional inter-QPU operations, with $d=2^{n_i}$.
We build results using up to $d=8$ Schmidt terms, beyond which the remaining Schmidt eigenvalues become negligible for the models that we consider. Correspondingly, we thus use $n_i=3$ inter-QPU gates. At each optimization step, if new parameters are added, we use $200$ random initial sets of parameters out of which the optimal ones are selected. 

\section{The interconnect advantage}
\begin{figure*}
    \begin{overpic}[width=0.8\textwidth]{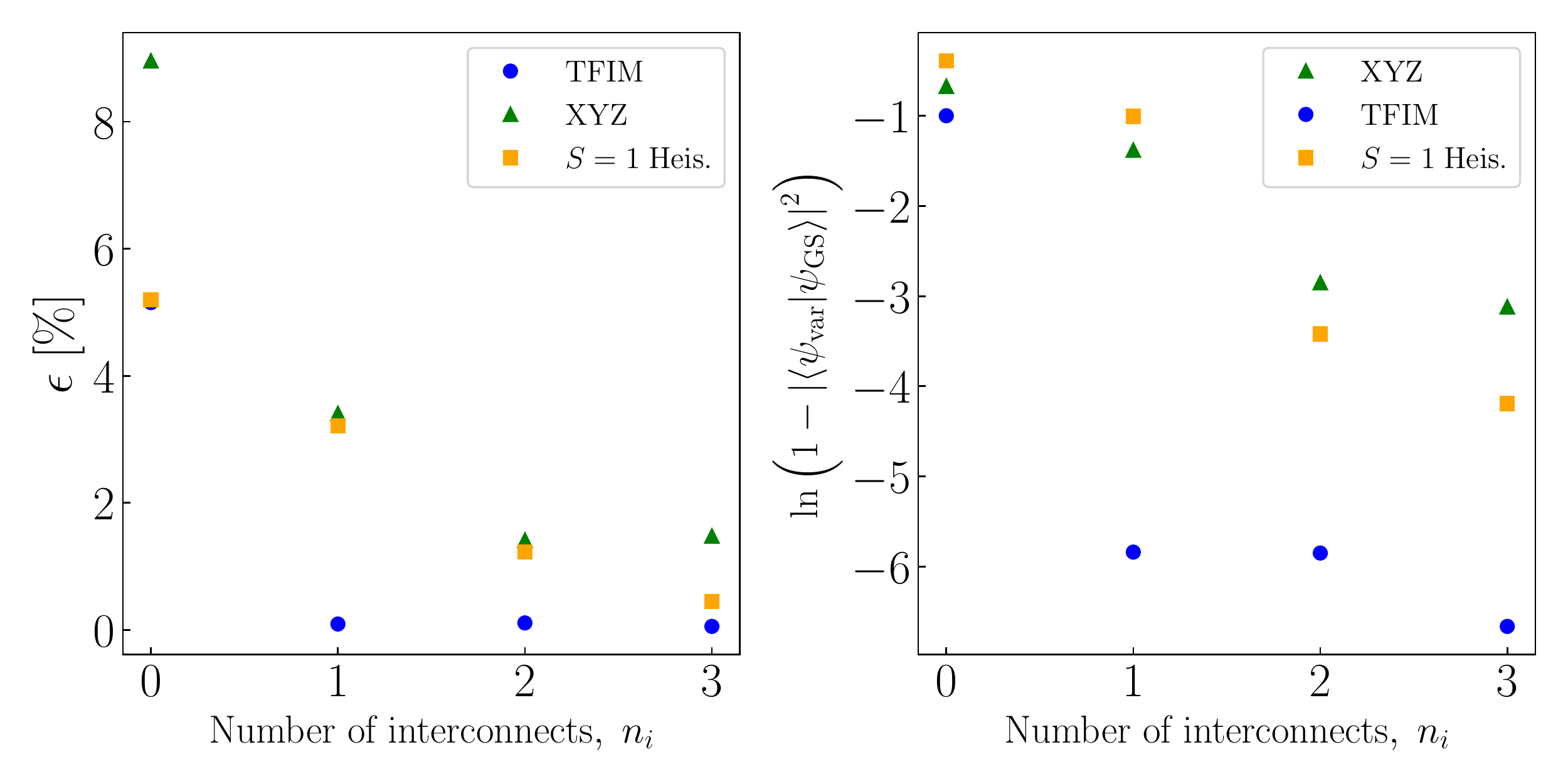} 
        \put(30,205){(a)}
        \put(240,205){(b)}
    \end{overpic}
    \caption{Comparison of (a) $\epsilon$, and (b) infidelity of the variational solution for the three models studied in the main text, given an increasing number of remote operations in a dual-quantum-core architecture of $N=12$ qubits. }
    \label{fig:ICnum}
\end{figure*}

Figure~\ref{fig:ICnum} shows the exponential benefit acquired with every interconnect use. As argued in the main text, each remote operation allows us to double the number of Schmidt terms. In Fig.~\ref{fig:ICnum} we display $\epsilon$ and the infidelities for the TFIM's least converging point and that of the anisotropic Heisenberg model (both discussed in the main text), and the $S=1$ Heisenberg model, all with $N=12$ qubits. The data suggests an improvement on both metrics as the number of remote operations increases. Extrapolating these quantities, or increasing $n_i$, might not continue that trend as the added Schmidt weight are decreasing, hence their contribution to both quantities diminishes. In addition, in practice the classical optimization task increases in complexity and one can expect a higher deviation from the theoretical infidelity (or energy) bound.
Nevertheless, the interconnect advantage is apparent.

\section{Ansatz expressiblity}
\begin{figure*}
    \begin{overpic}[width=0.8\textwidth]{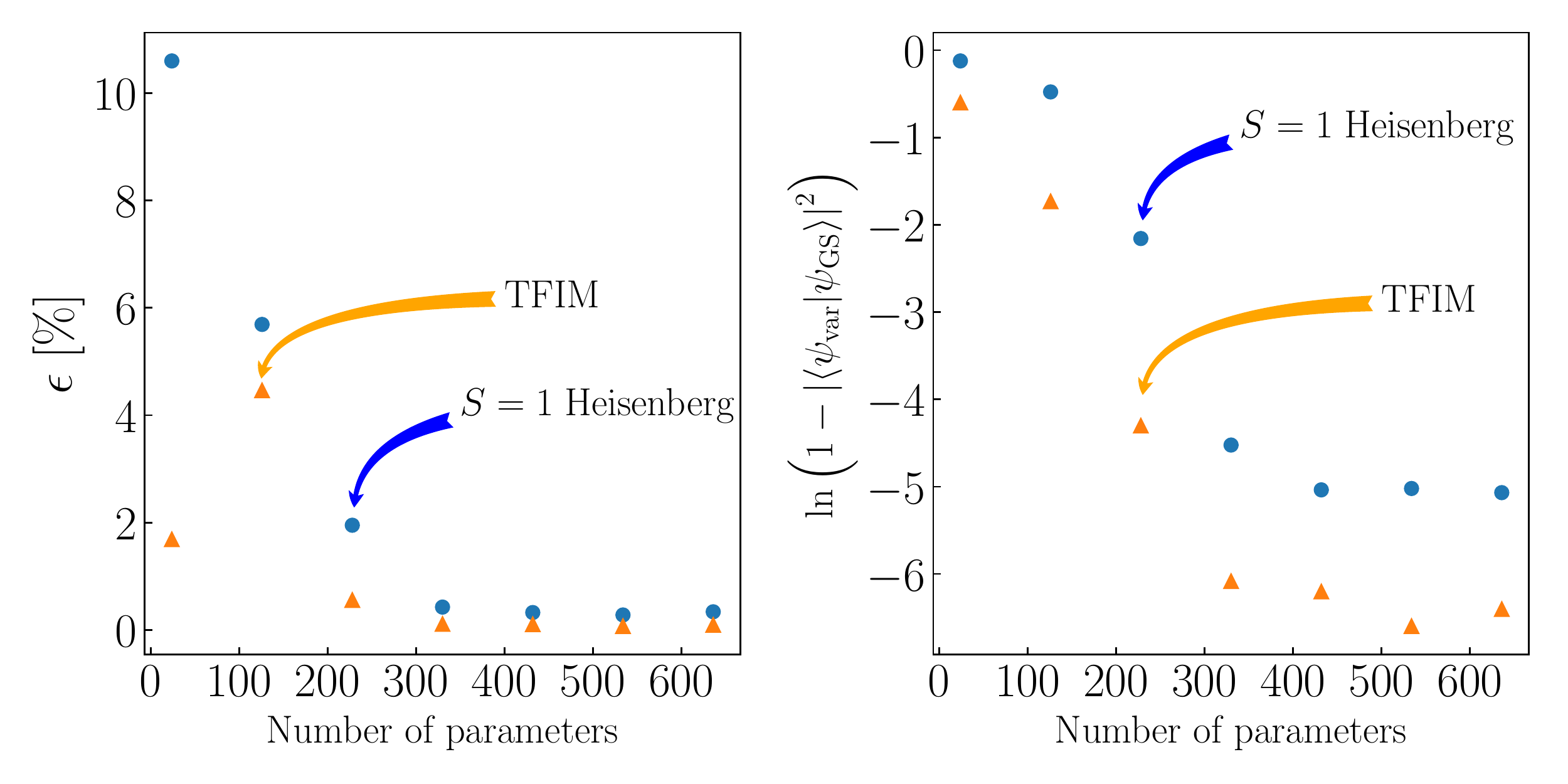} 
        \put(35,205){(a)}
        \put(240,205){(b)}
    \end{overpic}
    \caption{Comparing all-to-all connected ansätze performance for the $S=1$ Heisenberg and the TFIM models for $N=12$. 
    This comparison removes the expressibility question around the interconnected anzats, leaving only the classical optimization and the quality of the variational VQE ansatz for the specific model.  
    (a) The relative energy differences between the exact ground state and the variational ansatz, $\epsilon$, in percents. 
    (b) The log infidelity of the variational state. The x-axis depicts the number of variational parameters used, increments of this number correspond to adding another VQE layer (see panel (b) in Fig.~\ref{fig:VQEansatz}). 
    We use up to 7 layers ($636$ variational parameters) in comparison to 12 layers ($753$ parameters) in the interconnected case. Overall, the results of the TFIM model are more than an order of magnitude better compared to the $S=1$ Heisenberg model.}
    \label{fig:AKLTall2all}
\end{figure*}

To remove the effect of the interconnect itself we examine here a single QPU architecture with all-to-all connected qubits. The QPU is large enough to run the entire VQE instance, and no interconnect is necessary. The accuracy of this approach depends only on the expressibility of the specific ansatz and the capability of the classical optimizer. Results are shown in Fig.~\ref{fig:AKLTall2all}. We examine the $S=1$ Heisenberg model at $J(1,-1,0.5,1)$ and the TFIM at $h=0.73J$ (which are the least converging data points for $N=12$ in Figs.~\ref{fig:TFIM},~\ref{fig:Heis}), and increase the number of layers up to $m=7$. This amounts to $636$ variational parameters, as we wish to keep the number of parameters to be similar to the interconnected case. In Fig.~\ref{fig:AKLTall2all} (a) we show the relative energy difference between the variational energy and the exact ground state energy, while panel (b) shows the infidelity. With increasing number of layers---and the number of variational parameters as a consequence---we theoretically increase the expressibility of the circuit. However, we find that for many optimization attempts we see a saturation of the resulting energy around $m=5$ layers, which in fact then grows as one continue to increase $m$, probably due to limitations in the classical optimization. 
Nevertheless, comparing the resulting fidelities we find that the all-to-all architecture performs slightly better than the interconneced one, with a relative energy error of $0.08\%$ versus $0.09\%$ for the TFIM, and $0.3\%$ versus $1.4\%$ for the $S=1$ Heisenberg model. 

\section{Comparing architectures: all-to-all (single-core) and interconnected}
While the comparison throughout the text is to the separable solution, one could ask how does the \emph{limited} interconnected architecture performs compared to an all-to-all QPU of the same size. 

Starting with the TFIM and focusing on the least converging point at $h=0.73$, which was discussed in the main text (see Fig.~\ref{fig:TFIM}): The all-to-all architecture has an infidelity of $6.8 \cdot 10^{-4}$, compared with $9.1 \cdot 10^{-4}$ for the interconnected case. These fidelities corresponds to $\epsilon_{\rm all-to-all}=0.09\%$, and $\epsilon_{\rm interconnected}=0.08\%$ respectively.
As for the $S=1$ Heisenberg model, the infidelities are: $3.2\cdot 10^{-3}$ for the all-to-all and $7.6 \cdot 10^{-3}$ for the interconnected ansatz, and $\epsilon_{\rm all-to-all}=0.34\%$, and $\epsilon_{\rm interconnected}=0.33\%$ respectively. 
As one could expect, the variational energies do not correspond to fidelities and vise versa, but merely serve as a proxy. 

In conclusion, while the structure and expressibility of each of these ansätze are different, we can see only a slight advantage of the all-to-all  architecure. Though it has greater expressibility, it comes with a classical optimization cost as the number of vairational parameters scales as the number of participating qubits, $N^2$. This is what might limit the current performance of the single QPU  
architecture. However and more importantly, the real advantage of interconnected (multi-QPU) 
systems is the lower
complexity and overheads related to e.g., calibration, control, and noise reduction, challenges that are easier to handle in modular designs that utilize smaller processors.
\end{document}